\documentclass[conference]{IEEEtran}
\IEEEoverridecommandlockouts
\usepackage{cite}
\usepackage{amsmath,amssymb,amsfonts}
\usepackage{algorithm}
\usepackage{algorithmic}
\usepackage{subfigure}
\usepackage{caption}
\usepackage{graphicx}
\usepackage{bm}
\usepackage{textcomp}
\usepackage{xcolor}
\def\BibTeX{{\rm B\kern-.05em{\sc i\kern-.025em b}\kern-.08em
		T\kern-.1667em\lower.7ex\hbox{E}\kern-.125emX}}

\begin{document}

\title{ Age of Information Analysis for Multi-Priority Queue and NOMA Enabled C-V2X in IoV\\
	\thanks{This work was supported in part by the National Natural Science Foundation of China under Grant No. 61701197, in part by the National Key Research and Development Program of China under Grant No.2021YFA1000500(4), in part by the 111 Project under Grant No. B12018.}
}

\author{\IEEEauthorblockN{1\textsuperscript{st} Zheng Zhang}
	\IEEEauthorblockA{\textit{School of Internet of Things Engineering} \\
		\textit{Jiangnan University}\\
		Wuxi, China \\
		zhengzhang@stu.jiangnan.edu.cn}
	~\\
	\\
	
	\IEEEauthorblockN{3\textsuperscript{rd} Pingyi Fan}
	\IEEEauthorblockA{\textit{Department of Electronic Engineering} \\
		\textit{Tsinghua University}\\
		Beijing, China \\
		fpy@tsinghua.edu.cn}

	\and
	\IEEEauthorblockN{2\textsuperscript{nd} Qiong Wu*}
	\IEEEauthorblockA{\textit{School of Internet of Things Engineering} \\
		\textit{Jiangnan University}\\
		Wuxi, China \\
		qiongwu@jiangnan.edu.cn}
	Corresponding Author
	~\\
	\\
	
	\IEEEauthorblockN{4\textsuperscript{th} Ke Xiong}
	\IEEEauthorblockA{\textit{School of Computer and Information Technology} \\
		\textit{Beijing Jiaotong University}\\
		Beijing, China \\
		kxiong@bjtu.edu.cn}			
}
\maketitle

\begin{abstract}
	As development Internet-of-Vehicles (IoV) technology and demand for Intelligent Transportation Systems (ITS) increase, there is a growing need for real-time data and communication by vehicle users. Traditional request-based methods face challenges such as latency and bandwidth limitations. Mode 4 in Connected Vehicle-to-Everything (C-V2X) addresses latency and overhead issues through autonomous resource selection. However, Semi-Persistent Scheduling (SPS) based on distributed sensing may lead to increased collision. Non-Orthogonal Multiple Access (NOMA) can alleviate the problem of reduced packet reception probability due to collisions. Moreover, the concept of Age of Information (AoI) is introduced as a comprehensive metric reflecting reliability and latency performance, analyzing the impact of NOMA on C-V2X communication system. AoI indicates the time a message spends in both local waiting and transmission processes. In C-V2X, waiting process can be extended to queuing process, influenced by packet generation rate and Resource Reservation Interval (RRI). The transmission process is mainly affected by transmission delay and success rate. In C-V2X, a smaller selection window (SW) limits the number of available resources for vehicles, resulting in higher collision rates with increased number of vehicles. SW is generally equal to RRI, which not only affects AoI in  queuing process but also AoI in the transmission process. Therefore, this paper proposes an AoI estimation method based on multi-priority data type queues and considers the influence of NOMA on the AoI generated in both processes in C-V2X  system under different RRI conditions. This work aims to gain a better performance of C-V2X system comparing with some known algorithms..
\end{abstract}


\section{Introduction}
With the innovation and advancement of IoV technology, the development of ITS has catered to a range of in-vehicle applications such as automated navigation, collision warning, and multimedia entertainment \cite{FJ2010, QWHXL2019, QWJZ2014, QWJZ2014C, P2022ITS, wu2023ss}. Due to the high-speed mobility of vehicles, ensuring timely access for the required data and content through user requests is crucial \cite{QWSN2018, QWJZ2016, QWJZ2015P, Liuchen2021, wu2024wwh, wu2024urllc}. The traditional request-based approach involves users communicating with the base station, which then accesses the data center to retrieve the requested data\cite{wu2023YZ, wu2023XB, wu2023DX, wu2019SY, SW2021}. However, this method suffers from end-to-end delays and limited backhaul bandwidth \cite{Dai2019, KW2019, QW2018SYX, QW2024WSY, QW2015ICC, JFQW2010, PYF2002}. To address these issues, C-V2X proposes an interface called PC5 for communication between autonomous vehicles. PC5 offers two resource allocation methods: Mode 3, where User Equipment (UE) requests time and frequency-domain transmission resources from the eNodeB, and Mode 4, where UE autonomously selects resources without involving the cellular infrastructure. Mode 4 not only eliminates the limited coverage drawback but also minimizes interaction between base stations and vehicles, thereby resolving excessive delays and overhead \cite{Molina2017}.

In Mode 4, vehicles autonomously select communication resources using the SPS protocol based on distributed sensing. SPS allows vehicles to choose from several standardized message intervals (e.g., 10 Hz, 20 Hz, or 50 Hz). However, this increases the likelihood of collisions when multiple vehicles occupy the same resources with the same message transmission interval, leading to higher BLER \cite{3GPP2017}. NOMA is a potential solution for C-V2X communication, promising to enhance spectrum efficiency and handle large-scale vehicle communications, mitigating latency and packet reception probability degradation caused by high vehicle density \cite{Di2017, KX2014,XC2017}. Serial Interference Cancellation (SIC) is a well-known Multi-User Detection (MUD) technique used to extract overlapping signals, decoding different power levels of received signals occupying the same resource. High-power signals no longer interfere with other low-power signals after decoding, improving the Signal-to-Interference-plus-Noise Ratio (SINR) of low-power signals, reducing transmission failures due to collisions, and lowering BLER \cite{Zhang2017, JL, RJ, YG}.

Some studies have examined the effectiveness of NOMA applied in C-V2X. In \cite{Situ2020}, a NOMA receiver based on Continuous Interference Cancellation (SIC) and Joint Decoding (JD) was proposed, demonstrating its implementation in the current C-V2X communication and its ability to reduce BLER compared to traditional Orthogonal Multiple Access (OMA) methods. In \cite{Hirai2020}, TAKESHI et al. introduced SPS-NOMA based on UpLink Non-Orthogonal Multiple Access (ULNOMA), improving SIC's Signal-to-Noise Ratio (SNR) under broadcast scenarios and alleviating channel congestion. In \cite{Dey2022}, Utpal et al. proposed a model with a large-scale MIMO Jacobi detection algorithm for PHY layer of C-V2X, enhancing reliability by reducing bit error rate compared to existing PHY layer frameworks. These works demonstrated improvement of NOMA on reliability and transmission delay in the C-V2X system, where reliability and transmission delay were frequently used to assess vehicle communication performance in existing engineering and 3GPP standards. 

It is worth noting that these two metrics are often in trade-off, meaning an increase in reliability performance may come at the cost of increased delay. Therefore, a new metric is necessary to comprehensively reflect both reliability and latency performance, such as AoI, where lower average AoI indicates either lower latency with the same reliability or higher reliability with the same delay \cite{Kaul2012, ke2024,HZ}. In \cite{Peng2020}, Peng et al. adopted AoI to evaluate the MAC layer performance of C-V2X sidelink, proposing a Piggyback-based cooperative method for vehicles to inform each other of potential resource occupation, reducing collisions and exhibiting good AoI performance. In \cite{Mlika2022}, Zoubeir et al. presented a resource allocation problem based on NOMA, optimizing resource allocation to provide minimum AoI and high reliability for vehicle safety information. 

However, the aforementioned studies did not consider the AoI aspect of packet in the queue in C-V2X system. The packet received by the receiver describe the information at the time of packet generation by the transmitter, necessitating consideration of the queuing process. 
In \cite{Akar2021}, Akar et al. investigated the freshness of information in IoT-based state update systems using the AoI performance metric. They studied discrete-time servers in multi-source IoT systems, assuming Bernoulli arrivals of information packets and universally distributed discrete phase-type service times across all sources. Their analysis of AoI under various queuing disciplines was formulated in matrix-geometric terms.
In \cite{Zzx2021}, Zhang et al. considered a dual-server short-block wireless communication system to ensure real-time delivery of newly generated information at a relatively high update rate to its destination. Information is generated at a relatively high update rate, encoded into two short-block queues, and delivered in real-time through two parallel paths. The study based on the Markov-chain process investigated the AoI performance of the dual-queue system in the presence of block delivery errors.

The AoI of packet generated in the queue is influenced by packet generation rates and service rates, where the service rate in C-V2X depends on the RRI \cite{GPW2019}. However, for the same RRI, while the AoI of transmitter undergoes a short queuing time, the collision probability increases, leading to a potentially large AoI at the receiver \cite{MGM2019}. This motivates us to consider this work. In this paper, we propose an AoI calculation approach based on a multi-priority data type queue in C-V2X. We first design a queue model with four types of messages, each with different priorities, allowing the AoI of receiver to better reflect the ability to observe the  status of transmitter in a timely manner. We then consider the impact of NOMA on AoI in both processes, calculating AoI for different RRIs, and analyzing  effect of muti-priorities queues and NOMA on the C-V2X communication system.
\footnote{This paper has been submitted to WCSP. The source code has been released at: : https://github.com/qiongwu86/Analysis-of-the-Impact-of-Multi-Priority-Queue-and-NOMA-on-Age-of-Information-in-C-V2X}. The remaining parts of this article are as follows, Section 2 provides a brief introduction to the system model. Section 3 presents a description of the proposed modeling of approach in detail. In Section 4, we present some simulation results, followed by the conclusion in Section 5.

\section{System Model}

\begin{figure*}[!t]
	\centering
	\includegraphics[width=5.5in]{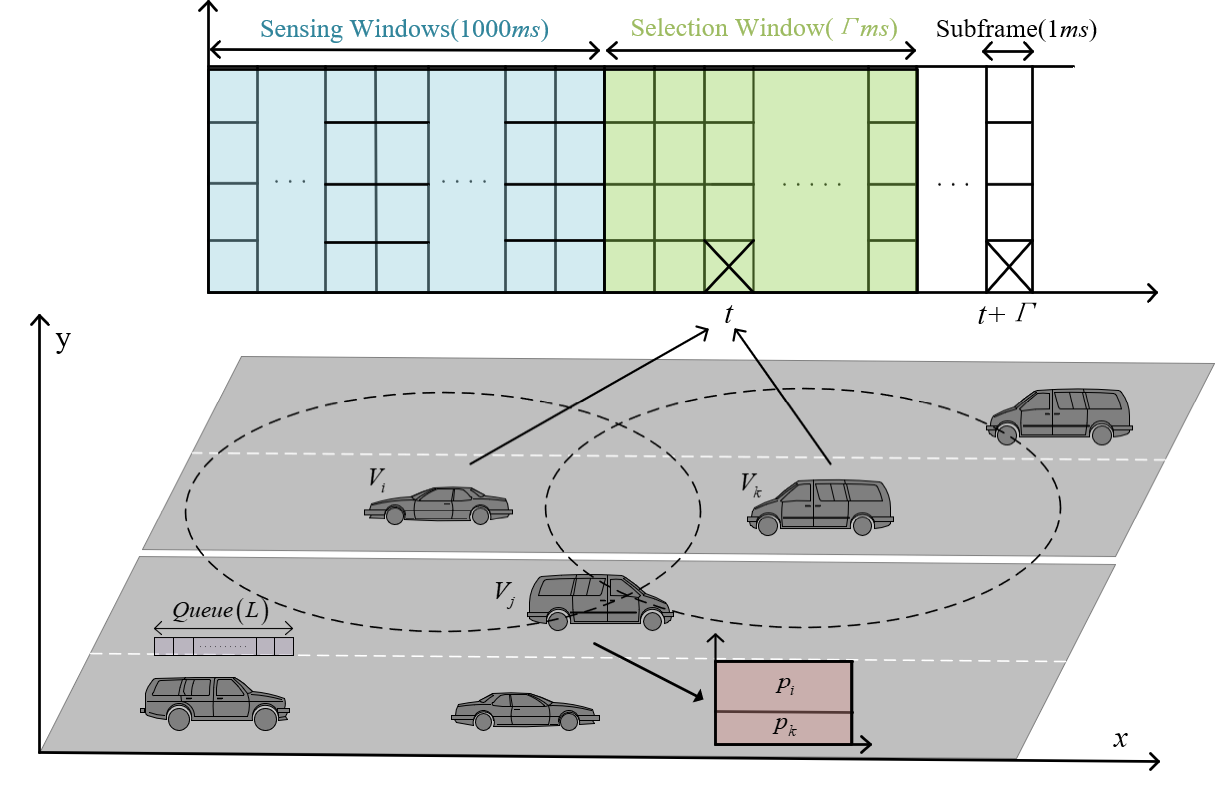}
	\caption{System Model}
	\label{fig:fig1}
\end{figure*}
We consider the system model as shown in Fig. \ref{fig:fig1}. In this model,  a cellular-based V2X communication system is with $N$ half-duplex vehicles. In C-V2X Mode 4, all communication resources are in a resource pool, with the smallest unit in the frequency dimension being an RB (Resource Block) with a size of 180 kHz. A sub-channel typically consists of 10 RBs. The time axis is set as a discrete value, with each time slot lasting 1 ms, representing the size of one sub-frame in the resource pool. In C-V2X Mode 4, there are four priority levels of message generating types. When the multi-priority queue is not empty, vehicles will reserve communication resources from the resource pool. The time interval between reserved resources is defined as the RRI , commonly taking values of 20 ms, 50 ms, or 100 ms. The initial value of reserved resources (RC) is defined as $500/RRI + \text{rand}(1000/RRI)$. Moreover, due to half-duplex communication, vehicles cannot receive signals while using communication resources.

We use AoI as the performance metric for this system and calculate the AoI generated in the multi-priority queue and the transmission process. Concerning the AoI in the queue, using communication resources will change the AoI and position of messages in the queue, while not using communication resources will keep them unchanged. Regarding the AoI in the transmission process, upon receiving a signal, the receiver updates its AoI for the transmitter to the AoI of the received signal; otherwise, it increments the AoI by one time slot. For simplicity, the receiver determines whether it receives the signal by comparing the SINR of the received signal with the SINR threshold as the criteria of successful transmission. When C-V2X employs OMA, it will compute SINR, where interferences mainly come from the collisions of other signals. For NOMA mode, SIC (see Eqn.(\ref{con:con10}) for detail) is used to improve the SINR of low-power signals in collisions, with SINR calculated for received signals in descending order of their power, where considering only signals with lower power than the current signal as interference. Additionally, in C-V2X, for the same RRI, the AoI in the queue may be very small while the receiver's AoI can be large. To better reflect the improvement of collisions by NOMA, it is necessary to separately calculate the AoI generated in the two processes and analyze the effect of NOMA  on the C-V2X communication system.

\section{Mathematical Model} 

In this section, we first establish a computational model to measure the AoI for a C-V2X system with a multi-priority message queue. Subsequently, we propose the use of NOMA based on SIC to improve the AoI in the C-V2X system.

\subsection{AoI in C-V2X}

1) C-V2X Queue Model

C-V2X Model 4 consists of four message types with the following priorities: HPD\textgreater DENM\textgreater CAM\textgreater MHD. CAM-type messages are generated periodically, while the other three types are triggered. The generation probability of a new CAM packet is $1/T_c$, where $T_c$ is the fixed packet generation period. The new packet generation probabilities for HPD, DENM, and MHD are represented by a Poisson distribution:

\begin{equation}
	P\left(\operatorname{arr}_{i, \mathrm{n}}^{t}=1\right)=\frac{\lambda_{\mathrm{n}}^{1}}{1 !} e^{-\lambda_{\mathrm{n}}}
	\label{con:con1}
\end{equation}
where $n \in \{\text{HPD, DENM, MHD}\}$, and $\lambda_n$ represents the number of packet arrivals for each type of message in a certain time period. For DENM and MHD, the new packets need to be retransmitted multiple times to ensure successful transmission, and this retransmission process is periodic. Thus, four corresponding First-In-First-Out (FIFO) queues are established based on different signal types. Their queue capacity is L and the queue length is $q$. When the queues meet $q\textless L$, corresponding new packets can be added and the same transmission opportunity that leaves the queue can be shared.

The transmission opportunity refers to vehicles can use their own reserved resources, defined as t. At this point, the vehicle transmits messages from different queues based on their priority. When the $q$ of the high-priority message queue is non-zero, the corresponding transmission action $s$ is set to 1, and for other lower-priority queues, $s$ is set to 0. Only when the length $q$ of the high-priority queue is zero, the $s$ of next lower-priority queue can be set to 1. A value of $s=1$ indicates that the messages in that queue can be transmitted; otherwise, they cannot be transmitted. Therefore, the expression for the transmission action $s$ of the multi-priority queue is:

\begin{equation}
\left\{\begin{array}{cc}s_{i, H}^{t}=1, s_{i, C}^{t}=0 & q_{i, H}^{t}=1 \\s_{i, H}^{t}=0, s_{i, C}^{t}=1 & q_{i, C}^{t}\left(1-q_{i, H}^{t}\right)=1 \\s_{i, H}^{t}=s_{i, C}^{t}=0 & \text { otherwise }\end{array}\right.
\label{con:con2}
\end{equation}

Due to the sequential prioritization, two priority queues can be used to represent the relationship between the $s$ of different queues. Let $s_{i, C}^{t}$ represent the transmission opportunity of the HPD queue at the time $t$, and $s_{i, H}^{t}$ represent the transmission opportunity of the CAM queue. When the $q$ of all queues are zero, it indicates that no packets can be transmitted, and at this moment, all $s$  are set to 0.

2) Multi-priority queue AoI Model

The cumulative AoI of messages in the queue is a result of the queuing process, defined as the time from packet generation to transmission. Thus, the processing rate affecting the queue length will also influence AoI. The AoI expression for each message in each queue is given by:

\begin{equation}
	\varphi_{i, n}^{t+1, b}=\left\{\begin{array}{cc}\varphi_{i, n}^{t, b+1}+1 & s_{i, n}^{t}=1 \\\varphi_{i, n}^{t, b}+1 & s_{i, n}^{t}=0\end{array}\right.
	\label{con:con3}
\end{equation}
where $n$ includes the all four queues, $b \in [1, q-1]$ represents the position of the message in the queue at time $t$ for vehicle $i$, and $\varphi_{n,i}^{t,b}$ denotes the AoI of the message in queue $n$ at position $b$ for vehicle $i$ at time $t$. When $s_{n,i}^t$=1, the position of all messages in the queue will move except for the first message; When $s_{n,i}^t$=0, all message positions in the queue remain unchanged. Here, $s=1$ represents the processing rate, and it depends on the 1/RRI in C-V2X. Thus, the RRI will influence the size of AoI.

The cumulative AoI during the communication process can be considered as the time spent to complete the transmission between the receiver and the transmitter, which is influenced by both the transmission delay and the transmission process. Additionally, since the received message reflects the situation at the time of packet generation by the transmitter, the receiver needs to inherit its size to represent the degree of understanding about the transmitter's situation. The AoI expression for receiver $j$ regarding transmitter $i$ is given by:

\begin{equation}
	\Phi_{i \rightarrow j}^{t+1}=\left\{\begin{array}{cc}\varphi_{i}^{t, 1}+1 & u_{i \rightarrow j}^{t}=1 \\\Phi_{i \rightarrow j}^{t}+1 & u_{i \rightarrow j}^{t}=0\end{array}\right.
	\label{con:con4}
\end{equation}
where $\Phi_{j,i}^t$ represents the AoI of vehicle $j$ regarding vehicle $i$ at time $t$, and $u_{j,i}^t$ indicates whether the message is successfully transmitted to $j$ by $i$. If $u_{j,i}^t$=1, $\Phi_{j,i}^t$ is equal to the $\varphi_{n,i}^{t,1}$ of the highest-priority message in the queue transmitted by vehicle $i$, plus the transmission delay. If $u_{j,i}^t$=0, $\Phi_{j,i}^{t+1}$ is equal to $\Phi_{j,i}^t$ plus the subframe size. According to Eq.(\ref{con:con3}) and (\ref{con:con4}), it can be observed that transmission failure will cause an greater increase in $\Phi_{j,i}^{t+1}$. In C-V2X model 4, the unique way of reserving resources by vehicles means that every transmission failure requires waiting for an RRI interval. The main reason for transmission failure is a low SINR caused by collisions, which can be addressed by NOMA.

\subsection{NOMA in C-V2X} 

1) Collisions in C-V2X

In C-V2X model 4, vehicles autonomously allocate resources and transmit data in a broadcast manner. When multiple vehicles reserve resources in the neighboring time slots, they may select the same resources. Moreover, due to the broadcast nature, multiple vehicles can simultaneously communicate with the same receiver. This situation leads to collisions when multiple vehicles use the same resources to communicate with the same receiver. According to \cite{Wijesiri2021}, the non collision probability in C-V2X model 4 can be expressed as:

\begin{equation}
	\begin{aligned}
		P_{\text{ncol}} & \approx \\
		& \left[1-\left[1-\prod_{i=0}^{\Gamma-1}\left(1-\frac{\pi}{1-\pi i}\right)\right] \frac{1-P_{\text{rk}}}{CSR-N_v+1}\right]^{N_v-1}
	\end{aligned}
	\label{con:con5}
\end{equation}
where $\pi$ represents the probability that a vehicle is in the moment of preparing to select a resource, refers to the probability of simultaneously satisfying the three following conditions: the vehicle queue is not empty, the RC is zero, and a new resource is being rescheduled. $P_{\text{rk}}$ represents the probability of a vehicle selecting a new resource when the reselection counter resets to zero. $CSR$ represents the total number of resources in the selection window. $N_v$ represents the total number of vehicles. $\Gamma$ is the size of SW, so $CSR$ follows $\Gamma$ Increase and increase.
 However, the variation range of $\Gamma$ is small, typically being 20, 50, or 100. Thus, when a specific value of $\Gamma$ is chosen, $\left[1-\prod_{i=0}^{\Gamma-1}\left(1-\frac{\pi}{1-\pi i}\right)\right] \frac{1-P_{r k}}{C S R-N_{v}+1}$ is a value that changes with the number of vehicles $N_v$, and as $N_v$ increases, $P_{\text{col}}$ also increases.Furthermore, changing the value of $\Gamma$ as the time interval for vehicles to transmit messages may reduce the collision probability, but it can also increase the waiting time of messages in the queue, leading to a larger AoI. Therefore, considering the impact of collisions, we explore the use of NOMA to mitigate this issue while changing $\Gamma$.

2) SINR Calculation

In C-V2X model 4, the size of transmission resources is not fixed. Vehicles calculate the required bandwidth B and the transmission rate threshold Rth for successful transmission within one subframe based on the size of the message Q, as expressed by:

\begin{equation}
	R_{\mathrm{th}}^{t}=B_{i} \log _{2}\left(1+\mathrm{SINR}_{\mathrm{th}}\right)
	\label{con:con6}
\end{equation}
where \(R_{\text{th}} = Q\) because the time taken to complete data transmission for a message must be less than the length of one subframe, i.e., \(Q/R \leq 1\). Next, the SINR threshold \(SINR_{\text{th}}\) is calculated based on the bandwidth and the transmission rate threshold, as shown in the following expression:

\begin{equation}
	\operatorname{SINR}_{\text {th }}=2^{Q_{i} / B_{i}}-1
	\label{con:con7}
\end{equation}

Due to the varying distances between vehicles at different time instants, the communication experiences different channel impairments, leading to varying communication rates at the receiver. Additionally, since multiple vehicles may use different communication resources within the same subframe, the receiver also receives signals at different rates within different communication resources. If the receiver receives only one signal in a resource, then the SINR between vehicle i and vehicle j at time t is given by:

\begin{equation}
	\operatorname{SINR}_{i \rightarrow j}^{t, n}=\frac{p_{i \rightarrow j}^{t}\left|h_{i \rightarrow j}^{t, n}\right|^{2}}{\sigma^{2}}
	\label{con:con8}
\end{equation}
where $P_i$ is the transmission power of vehicle i, $\left|h_{i \rightarrow j}^{t, n}\right|^{2}$ is the channel gain between vehicles i and j for resource n, and $N_0$ is the noise energy. However, due to the half-duplex resource selection scheme in C-V2X, vehicles may select the same resource block when choosing resources, leading to collisions when they use the same resource for transmission. In this case, the wireless information transmission rate is defined as:

\begin{equation}
	\operatorname{SINR}_{i \rightarrow r}^{t, n}=\frac{p_{i \rightarrow j}^{t}\left|h_{i \rightarrow j}^{t, n}\right|^{2}}{\sum_{m \in N_{m}} p_{m \rightarrow j}^{t}\left|h_{m \rightarrow j}^{t, n}\right|^{2}+\sigma^{2}}
	\label{con:con9}
\end{equation}
where $m$ represents interfering vehicles in the C-V2X scenario, and their transmission energy is denoted as $P_m$. It can be observed that when multiple vehicles use the same resource, the transmission rate decreases, and if the transmission rate is too low, it may result in transmission failure.

Therefore, to address this issue, we introduce NOMA based on SIC. The receiver sorts the received signals in descending order of received power, considering the maximum received power signal as the target signal and the rest as interference signals. After decoding and removing the target signal, this process is repeated for all signals to compute their SINR. Let $N_k = \{k\in N\setminus i | {p_{i \rightarrow j}^{t}\left|h_{i \rightarrow j}^{t, n}\right|^{2}} > {p_{k \rightarrow j}^{t}\left|h_{k \rightarrow j}^{t, n}\right|^{2}}\}$ represent the set of vehicles whose received power is weaker than vehicle i. Then, the SINR for vehicle i is given by:

\begin{equation}
	\operatorname{SINR}_{i \rightarrow r}^{t, n}=\frac{p_{i \rightarrow j}^{t}\left|h_{i \rightarrow j}^{t, n}\right|^{2}}{\sum_{k \in N_{k}} p_{k \rightarrow j}^{t}\left|h_{k \rightarrow j}^{t, n}\right|^{2}+\sigma^{2}}
	\label{con:con10}
\end{equation}

It can be seen that using NOMA in the event of a collision can reduce the interference on the target signal and increase SINR, thereby reducing the possibility of SINR being lower than the threshold and leading to transmission failure. It is generally believed that $\left|h_{i \rightarrow j}^{t, n}\right|^{2}= c_{ij}d_{ij}$, $c_{ij}$ represents the coefficient between vehicle i and vehicle j, $d_{ij}$ denotes the distance between vehicle i and vehicle j, and $\eta$ is the path loss exponent. The value of $d_{ij}$ depends on the positions of vehicles i and j at time t.

3) Vehicle Movement
Consider a two-way road with a length of \(D\) and \(U/2\) lanes in each direction. We establish a coordinate system where the position of vehicle \(i\) at time \(t\) is defined as \((x_i^t, y_i^u)\). The origin of the coordinate system is set at the leftmost position of the road, with the \(x\)-axis representing the direction of vehicle movement, and the \(y\)-axis representing different lanes. Assuming that the vehicle updates its position every very short time, the speed \(V\) of vehicle \(i\) can be considered constant:

\begin{equation}
	x_{i}^{t+1}=x_{i}^{t}+\delta V \tau, x_{i}^{t} \in[0, D]
	\label{con:con11}
\end{equation}
where \(\delta\) represents the direction of vehicle, and set $x_{i}^{0} \in[0, D]$ be the initial position of the vehicle. And $y_i^u$ depends on the lane index $u \in [1, \ldots, U] $ of vehicle $i$, calculated as:

\begin{equation}
	y_{i}^{u}=u d_{y}-y_{0}
	\label{con:con12}
\end{equation}
where \(d_y\) is the width of one lane, and \(y_0\) represents the distance of vehicle \(i\) to the edge of the lane. Typically, \(y_0\) is taken as \(1/2 d_y\), which is half of the lane width.

\section{Numerical Results and Discussion}

Simulation Tool is MATLAB R2021b. The simulation is conducted based on the existing work with additional modifications. In this section, we present the simulation results for the comparison of AoI in C-V2X under NOMA and OMA, validating the impact of NOMA on AoI in C-V2X.

\subsection{Simulation Settings}
According to the C-V2X standard, we use a 10 MHz bandwidth with a total of 50 RBs. The message size is set to 500 Bytes, and QPSK modulation is applied for propagation. The $T_C$ can be an integer multiple of 100 ms between 100 ms and 1 s, typically set to 100 ms. The arrival rates $\lambda_H$, $\lambda_D$, and $\lambda_M$ for different message types are set to 0.0001. The $T_H$ and $T_D$ are set to 100 ms and 500 ms, respectively. The $K_H$ and $K_D$ are set to 8 and 5, respectively. The transmission power of all vehicles is uniformly set to 23 dBm, and the speed is set to 120 km/h.

\subsection{Performance Evaluation}

Take the mean of $\varphi_{i, n}^{t, b}$ to represent the total AoI of packets in the queue of each vehicle on average, reflecting the queuing situation.

\begin{equation}
	\overline{\varphi}=\frac{1}{N_{v}}\frac{1}{N}\frac{1}{b}\sum_{i=1}^{N_{v}} \sum_{n \in N} \sum_{b}^{{L}} \varphi_{i, n}^{t, b} 
	\label{con:con13}
\end{equation}
where $Nv$ represents the total number of vehicles in the scene, and $N$ is 4, indicating four types of messages. The $\Delta^{t}$ is defined as the average of $\Phi_{i \rightarrow j}^{t}$ between each pair of vehicles.

\begin{equation}
\Delta^{t}=\frac{1}{N_{v}} \frac{1}{N_{v}-1} \sum_{j=1}^{N_{v}} \sum_{i=1}^{N_{v}-1} \Phi_{i \rightarrow j}^{t}
\label{con:con14}
\end{equation}
where, $\Phi_{i \rightarrow j}^{t}$ represents the AoI between vehicles $i$ and $j$ at time $t$, which have already computed the AoI between vehicles acting as transmitters and receivers at each moment, taking into consideration the impact of half-duplex communication. Thus, the summation involves calculating the mean AoI between different vehicle pairs $i$ and $j$. In general, a higher $\Delta^{t}$ indicates a higher number of transmission failures.

\begin{table*}[ht]
	\centering
	\caption{Communication success rate in C-V2X}
	\begin{tabular}{|c|c|c|c|c|c|c|}
		\hline
		$N_v$ & \multicolumn{3}{|c|}{30} & \multicolumn{3}{|c|}{50}\\
		\hline
		RRI & 20 & 50 & 100 & 20 & 50 & 100\\
		\hline
		OMA & 0.82891 &	0.83738 & 0.91560 &	0.75367 & 0.80050 &	0.85184 \\
		\hline
		NOMA & 0.89488 & 0.93332 & 0.97274 & 0.87902 & 0.92636 & 0.95356 \\
		\hline
	\end{tabular}
	\label{tab1}
\end{table*}

Table \ref{tab1} presents the communication success rate for different vehicle counts under different values of $\Gamma$, which represents the proportion of successfully received messages compared to all received messages. The vehicle density is set at 60 and 100 vehicles\slash km, and the length of load is set at 500 m. So it can be observed that the success rate is generally higher for 30 vehicles compared to 50 vehicles. This is because an increase in the number of vehicles enhances the probability of reserving resources at the same time, leading to a higher likelihood of resource contention.
Furthermore, the success rate at $\Gamma = 20$ ms is consistently lowest. This is because a smaller CSR will resulting in a higher probability of each resource being reserved by multiple vehicles in scenarios with a greater number of vehicles. Moreover, it is worth noting that, for a given number of vehicles, regardless of the value of $\Gamma$, employing NOMA technology consistently yields better transmission success rates compared to C-V2X with OMA.

\begin{figure}[!t]
	\centering
	\includegraphics[width=3.5in]{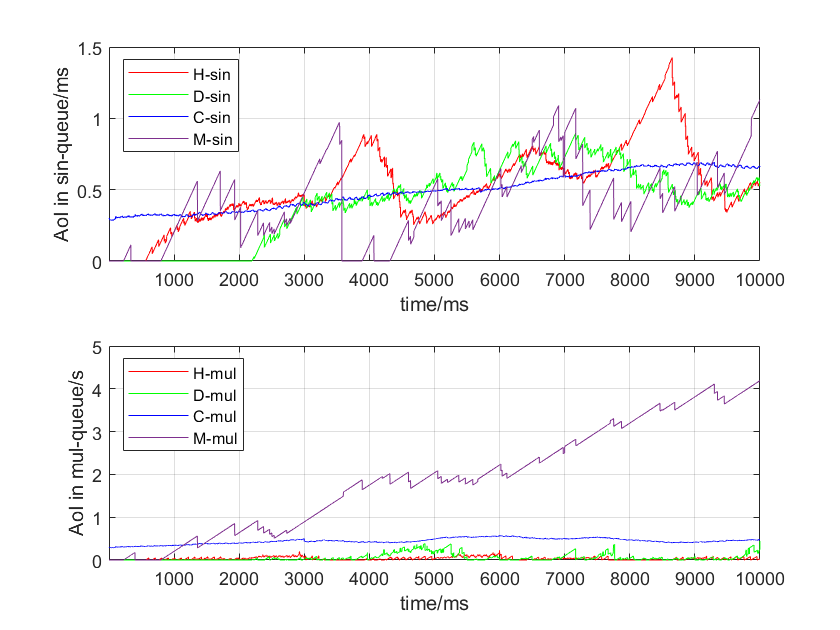}
	\caption{AvgAoI in different Queues}
	\label{smtr}
\end{figure}
Fig. \ref{smtr} shows the age variation trend of various messages when the number of vehicles is set to 50 and $\Gamma$ is set to 100 ms. The upper part of the graph shows the age change of information when a vehicle uses a single queue, while the lower part shows the situation when a vehicle uses a multi priority queue. At this point, the messages in the queue will pile up, leading to an increasing AoI. 
It can be seen that when there is only one queue, multiple messages queue together, so their overall trend of AoI change is similar.
In a multi priority queue, although the probability of generating MHD messages is the lowest, its AoI is the highest due to its lowest priority. So in multi priority queues, the AoI of high priority messages decreases to a certain extent, ensuring the timeliness of high priority messages.
\begin{figure}[!t]
	\centering
	\includegraphics[width=3.5in]{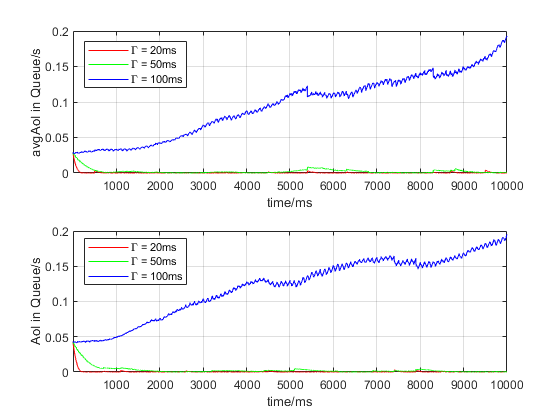}
	\caption{AvgAoI in Queues}
	\label{queuetrend}
\end{figure}

Fig. \ref{queuetrend} presents the $\overline{\varphi}$ with different values of $\Gamma$ when the number of vehicles is 30 and 50, respectively. When $\Gamma$ is 20 ms, the comprehensive production cycle of various types of packets is close to 100 ms (much larger than $\Gamma$), so the queue remains empty for most of the time, the $\varphi_{i, n}^{t, b}$ hardly increases. Similarly, when $\Gamma$ is 50 ms, the $\overline{\varphi}$ is also relatively small. However, when $\Gamma$ is 100 ms, which is greater than the production cycle, packets experience more time while waiting in the queue. In addition, it can be observed that when there are more vehicles, the value of $\overline{\varphi}$ increases to a certain extent compared to when there are fewer vehicles.

\begin{figure}[!t]
	\centering
	\includegraphics[width=3.5in]{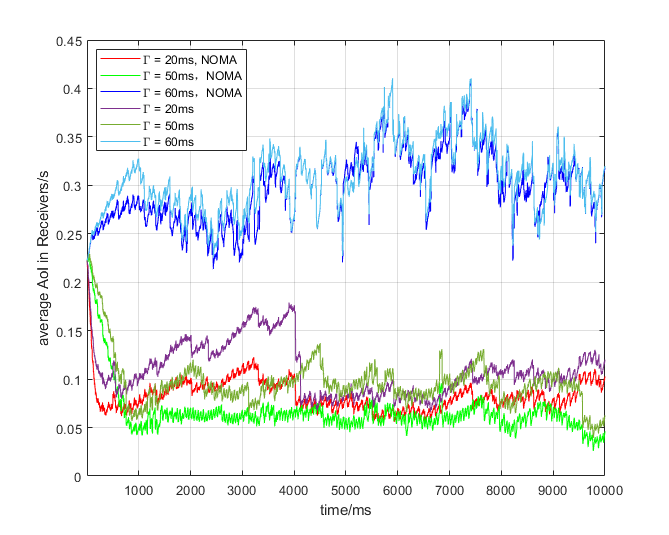}
	\caption{$N_v$ = 30}
	\label{r60}
\end{figure}
Fig. \ref{r60} shows the $\Delta^{t}$ of 30 vehicles at different $\Gamma$. It can be observed that for different $\Gamma$, NOMA consistently decreases $\Delta^{t}$. As shown in Table \ref {tab1}, when $\Gamma$ is 20ms, the collision probability is relatively high, resulting in the message spending the shorter time in the queue, but the information age of the receiving end is not being smaller. By contrast, when $\Gamma$ is 100 ms, the collision probability is lower. However, due to the accumulation of $\varphi_ {i, n}^{t, b}$, it continues to rise and instead becomes the oldest information age among the three situations. When $\Gamma$ is 50ms, the collision rate is the lowest among the three considered cases, and $\varphi_{i, n}^{t, b}$ does not accumulate, resulting in the lowest $\Delta^{t}$ throughout the entire observation period.

\begin{figure}[!t]
	\centering
	\includegraphics[width=3.5in]{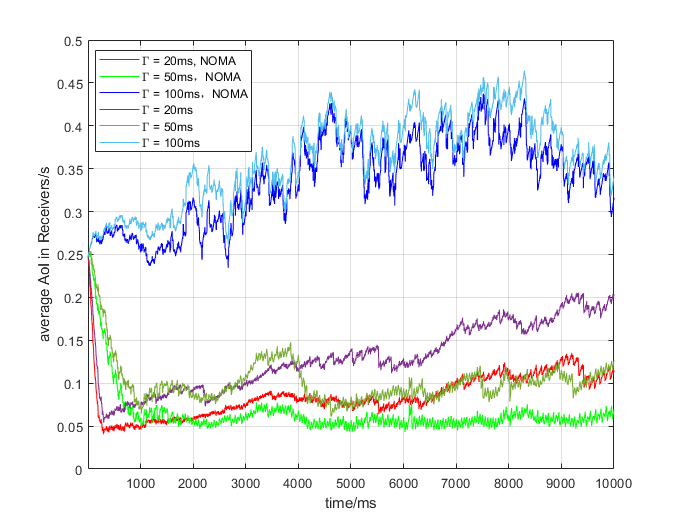}
	\caption{$N_v$ = 50}
	\label{r100}
\end{figure}

Fig. \ref{r100} depicts the $\Delta^{t}$  for 100 vehicles under different values of $\Gamma$. A comparison between Fig. \ref{r60} and Fig. \ref{r100} reveals that an increase in the number of vehicles leads to a general upward trend for all cases, which is attributed to the rising collision probability with an increasing number of vehicles. Additionally, the trends in these two cases are similar to those shown in Fig. \ref{r60}, with the highest information age at $\Gamma$=100ms and the lowest information age at $\Gamma$=50ms. NOMA has positive effects on information age for different $\Gamma$.

\section{conclusion}
This paper considers the mobility of vehicles and analyzes the impact of NOMA on the AoI in different scenarios. Firstly, we propose a novel multi-priority queue AoI calculation model for C-V2X communication. Then, we investigate the improvement in transmission success rate using NOMA based on SIC to observe changes in AoI. The following conclusions are drawn from our analysis:
\begin{itemize}
	\item Under the same $\Gamma$, the collision probability varies with different vehicle counts. Adjusting $\Gamma$ can reduce collision occurrences, but it must be carefully selected to avoid increased AoI. 
	\item NOMA enhances SINR and reduces the effects of collisions, leading to decreased AoI in various scenarios in C-V2X system.
\end{itemize}

\end{document}